\documentclass[]{raa}

\usepackage{graphicx,times}
\usepackage{amssymb,amsmath}
\usepackage[a4paper]{geometry}

\begin{document}

   \title{Prospects for detecting ultra-high-energy particles with FAST}
   \volnopage{Vol.0 (200x) No.0, 000--000}      
   \setcounter{page}{1}          

   \author{
   C.W. James
   	\inst{1}
    \and J.D. Bray
	\inst{2}
   \and R.D. Ekers
   	\inst{3}
   }

   \institute{ECAP, Univ.\ of Erlangen-N\"urnberg, Erwin-Rommel-Str. 1, 91058 Erlangen, Germany; {\it clancy.james@physik.uni-erlangen.de}\\
         \and
             School of Physics \& Astronomy, Univ.\ of Southampton, SO17 1BJ, United Kingdom\\
        \and
             CSIRO Astronomy \& Space Science, PO Box 76, Epping, NSW 1710, Australia\\
   }

   \date{Received~~2009 month day; accepted~~2009~~month day}

\abstract{
The origin of the highest-energy particles in nature, the ultra-high-energy (UHE) cosmic rays, is still unknown. In order to resolve this mystery, very large detectors are required to probe the low flux of these particles - or to detect the as-yet unobserved flux of UHE neutrinos predicted from their interactions. The `lunar Askaryan technique' is a method to do both. When energetic particles interact in a dense medium, the Askaryan effect produces intense coherent pulses of radiation in the MHz--GHz range. By using radio telescopes to observe the Moon and look for nanosecond pulses, the entire visible lunar surface ($20$ million km$^2$) can be used as an UHE particle detector. A large effective area over a broad bandwidth is the primary telescope requirement for lunar observations, which makes large single-aperture instruments such as the Five-Hundred-Meter Aperture Spherical Radio Telescope (FAST) well-suited to the technique. In this contribution, we describe the lunar Askaryan technique and its unique observational requirements. Estimates of the sensitivity of FAST to both the UHE cosmic ray and neutrino flux are given, and we describe the methods by which lunar observations with FAST, particularly if equipped with a broadband phased-array feed, could detect the flux of UHE cosmic rays.
}

   \authorrunning{C.W. James, J.D. Bray, and R.D. Ekers} 
   \titlerunning{ UHE cosmic ray detection with FAST } 

   \maketitle

\section{Introduction}           
\label{sect:intro}

The ultra-high-energy (UHE) cosmic rays are the highest-energy particles in nature, with a flux extending up to at least $100$ EeV ($10^{20}$~eV). Their origin remains unknown, largely due to the bending of their trajectories in cosmic magnetic fields. Only at extreme energies of $5.6 \cdot 10^{19}$~eV and above have these particles been shown to retain some directional information, as shown by the correlation of the arrival directions of these particles with nearby AGN (\cite{AugerScience07}). However, the AGN used for the correlation mostly serve as proxies for the large-scale matter distribution in the local universe, and the correlation remains weak (\cite{Auger2010}). The reason why the source(s) of the UHE cosmic rays within this large-scale distribution cannot be identified is simple: in this energy range, the flux is approximately one particle per square kilometre per century, so that even the $3000$~km$^2$ Pierre Auger observatory in Argentina only detects of order $30$ particles above $5.6 \cdot 10^{19}$~eV per year. In order to determine the origin of these particles, an even larger detector is required. However, the cost of building such a detector in the style of the Pierre Auger observatory has so-far proved too great, motivating the use of new techniques.

Another method to determine the cosmic-ray origin is to detect the neutrinos produced from cosmic-ray interactions with background photon fields, and/or any flux expected from their interactions during acceleration (\cite{Berezinskij69}). Since neutrinos are uncharged and weakly interacting, any high-energy neutrino flux will arrive at Earth directly from the source, so that detecting this flux should point back to the source of the UHE cosmic rays, thus giving another method to resolve the UHE cosmic-ray mystery.

The lunar Askaryan technique (\cite{Dagkesamanskii89}) is a method to detect both the highest-energy cosmic rays and neutrinos, using the Askaryan effect (\cite{Askaryan62,Askaryan65}). Askaryan predicted that a particle cascade in a medium would develop an excess of negative charge due to the entrainment of atomic electrons, and the annihilation of positrons in-flight. This excess charge --- of order $10$\% of the total charge --- will radiate coherently at wavelengths comparable to the dimensions of the cascade. Thus, the total power radiated will scale with the square of the excess charge, and hence with the square of the primary particle energy. Experiments utilising the Askaryan effect therefore tend to target the highest-energy particles. Cosmic rays impacting the Moon will produce cascades approximately $10$~cm wide and a few metres long (\cite{Alvarez-Muniz06}), so that coherent radiation is expected in the $100$~MHz to a few GHz regime. By observing the Moon with a ground-based radio telescope, its entire visible surface of twenty million square kilometres can be used as a cosmic ray detector.

The Askaryan effect has since been confirmed via accelerator experiments at SLAC (\cite{Saltzberg01}), and numerous experiments have utilised the lunar Askaryan technique to search for UHE particles impacting the Moon. These have covered observations at Parkes and the ATCA by the LUNASKA collaboration (\cite{Hankins96,James10}); GLUE at Goldstone (\cite{Gorham04}); a series of observations using the telescope at Kalyazin (\cite{Beresnyak05}); the NuMoon project's observations with the WSRT (\cite{Buitink10});
and RESUN at the VLA/EVLA (\cite{Jaeger10}). While these experiments made significant advances in the unique techniques required to detect Askaryan radiation, none had the sensitivity to identify the signature pulses produced by such particle interactions. This is expected to change with the Five-Hundred-Meter Aperture Spherical Radio Telescope (FAST), which offers the first opportunity to detect the existing cosmic ray flux with the lunar Askaryan technique.

In Section \ref{sect:emission}, the radio emission from particle cascades in the Moon is described, outlining the basic properties of the signal produced by UHE cosmic ray and neutrino interactions. The unique observing mode required to detect such a signal is described in Section \ref{sect:observations}, focussing in particular on dedispersion, RFI discrimination, and triggering. The expected sensitivity of FAST to UHE cosmic rays and neutrinos using the lunar Askaryan technique is simulated in Section \ref{sect:simulations}, using a detailed Monte Carlo code developed for the LUNASKA experiments at Parkes and the ATCA. Using these results, we discuss the potential of the lunar Askaryan technique with FAST in regards to UHE particle searches in Section \ref{sect:discussion}.

\section{Radio Emission from UHE Particles Impacting the Moon}
\label{sect:emission}

When a high-energy particle interacts in a medium, it produces a cascade of secondaries with initially balanced positive and negative charges. Compton, Moeller, and Bhabha scattering of medium electrons (by gamma rays, electrons, and positrons, respectively) will entrain medium electrons into the cascade, while occasionally cascade positrons will annihilate in-flight. As predicted by G. A. Askaryan, these processes produce a total negative charge excess of order $10$\%. An observer will see this charge excess rapidly appear and disappear as a sudden `flash', producing a pulse of radiation lasting no longer than the duration of the cascade as seen by an observer. This `Askaryan radiation' will therefore be coherent at wavelengths greater than the apparent size of the cascade. At the Cherenkov angle ($\theta_C = \cos^{-1} n^{-1}$, where $n$ is the refractive index), emission along the entire length of the cascade will radiate coherently, and coherency will be limited only by the width of the cascade\footnote{Although the peak emission is at the Cherenkov angle, the radiation in the Askaryan effect is not primarily Cherenkov radiation --- see the appendix of \cite{Endpoints} for a discussion.}.  At angles far from $\theta_C$ however, high-frequency radiation along the length of the cascade will tend to destructively interfere, and coherency will be observed only at lower frequencies.

\begin{figure}
   \centering
   \includegraphics[width=0.5\textwidth, angle=0]{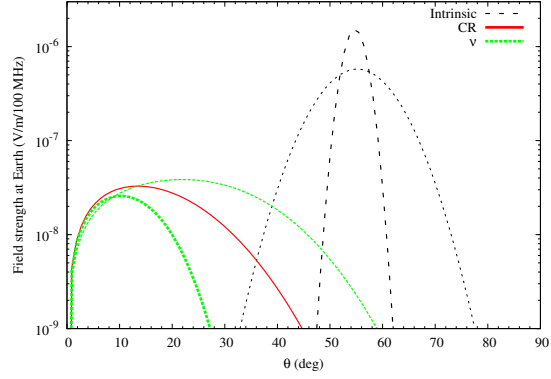}
   \caption{Spectral radiation strength at Earth from $10^{20}$~eV particle interactions in the lunar regolith, at $1$~GHz (thick) and $300$~MHz (thin), as a function of angle to the lunar surface. Black `intrinsic': emission from a cascade parallel to the surface (direction $\theta=0$), ignoring absorption in, and refraction and transmission through, the lunar surface. Note the peak at the Cherenkov angle of $55^{\circ}$. Red `CR': characteristic emission of a cosmic ray directed $10^{\circ}$ into the lunar surface --- the $1$~GHz emission is not visible on this scale. Green `$\nu$': characteristic emission of a neutrino parallel to the surface, at depth of $10$~m, giving only $20$\% of its energy to a particle cascade.}
   \label{fig:sig_strength}
\end{figure}

\begin{figure}
   \centering
   \includegraphics[width=0.45\textwidth, angle=0]{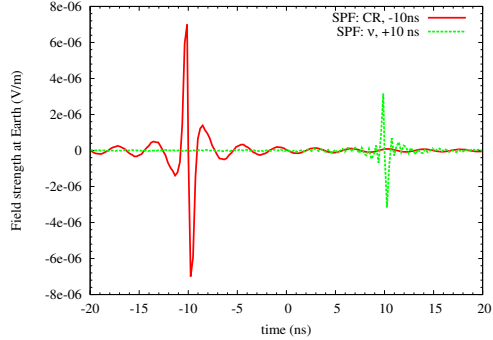}
   \caption{Time-domain signals of the characteristic cosmic ray (CR: red) and neutrino ($\nu$: green dotted) events shown in Fig.~\ref{fig:sig_strength} as viewed from Earth. The assumed emission angle is $5^{\circ}$, and the bandwidth $270$~MHz to $1.45$~GHz.}
   \label{fig:time_domain}
\end{figure}

Cosmic rays impacting the Moon will produce hadronic cascades in the lunar regolith, the outer-most layer of sand-like material covering the Moon. This material has an approximate density of $\rho \sim 1.8$~g/cm$^3$ (\cite{OlhoeftStrangway75}), producing cascades which are $\sim10$~cm wide and a few metres long (\cite{Alvarez-Muniz06}). Therefore, emission from lunar cascades can peak in the GHz regime. The expected emission from a $10^{20}$~eV hadronic cascade in the lunar regolith at $300$~MHz and $1$~GHz is shown by the black `intrinsic' lines in Fig.~\ref{fig:sig_strength}, as a function of the angle from the shower axis (for more details of signal simulation, see Sec.~\ref{sect:simulations}). The $1$~GHz emission is stronger, but the $300$~MHz emission broader.

The emission that escapes the regolith to be seen by an observer however is strongly dependent on the interaction geometry. Since the Cherenkov angle is the complement of the angle of total internal reflection, the peak emission from cascades parallel to the surface will not escape the regolith. This is shown by the green `$\nu$' lines in Fig.~\ref{fig:sig_strength}. Since cosmic rays will always interact pointing into the surface, the suppression due to transmission effects is even greater, as shown by the red `CR' lines in Fig.~\ref{fig:sig_strength}. At high frequencies therefore, only those cosmic rays impacting the surface at low incident angles will be visible. These transmission effects also mean that an observer will preferentially observe radiation coming from the lunar limb, especially at higher frequencies --- this is often termed the `limb brightening' effect.

The Askaryan radiation emitted from particle cascades is always linearly polarised in the plane of the observer and the shower axis. Radiation escaping the surface will thus tend to be polarised parallel to the local surface normal. Combined with the limb-brightening effect, Askaryan pulses from UHE particles hitting the Moon will tend to appear as linearly-polarised, bandwidth-limited impulses with polarisation pointing radially outward from the Moon.

The emission from UHE neutrino-induced cascades will appear similarly to that from cosmic-ray-induced cascades, with a few important differences. Firstly, neutrinos will only deposit of order $20$\% of their energy as hadronic cascades. In neutral-current (NC) interactions, the remaining $80$\% will be carried away by the neutrino, while the lepton produced in charged-current (CC) interactions will deposit its energy over a large distance, producing almost no visible emission. Unlike cosmic rays, neutrinos can penetrate a large fraction of the Moon before interacting (interaction length of $216$~km water equivalent at $10^{20}$~eV; \cite{Gandhi98}). While they will not pass through the entire Moon, they can `skim' the outer layers, and interact near the surface so that the resulting cascade points upwards, allowing high-frequency emission to escape. However, the majority of neutrinos will interact too deeply in the Moon to produce visible radiation. The absorption length is such that the emitted field strength $E$ at frequency $f$ reduces approximately as: $E \sim \exp\{ -(d/18~{\rm m}) (f/ 1~{\rm GHz})\}$ for an interaction at depth $d$, so that the interaction volume accessible at low frequencies is greater. Note that while the `lunar regolith' (of $2$--$10$ m thickness; \cite{Shkuratov01}) is often used as shorthand for the lunar material with which UHE particles will interact, in fact all lunar rock should be relatively radio-transparent, so the depth of the regolith layer will not limit the interaction volume of UHE neutrinos.

\begin{figure}
   \centering
   \includegraphics[width=0.45 \textwidth, angle=0]{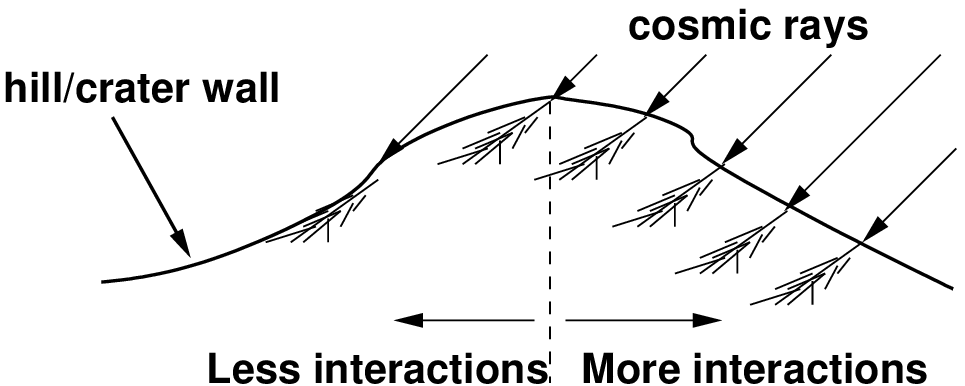} \includegraphics[width=0.35\textwidth, angle=0]{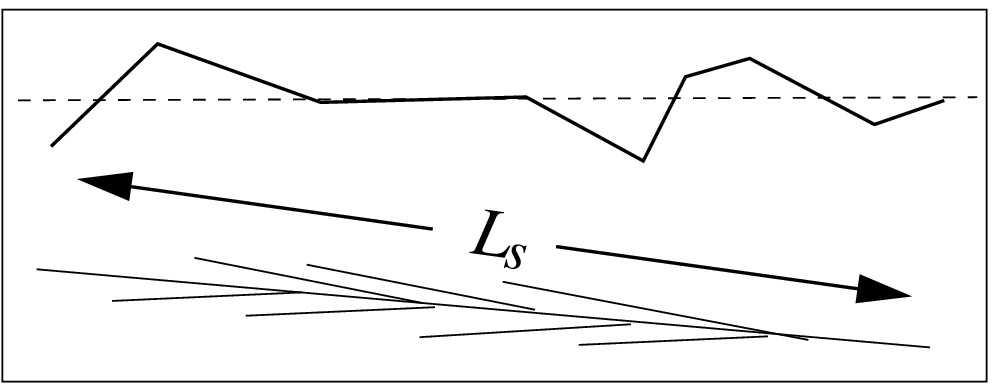}
   \caption{Sketches indicating the effects of lunar surface roughness. Left: cosmic rays will tend to interact on backwards-sloping surface features, making the escape of radiation more difficult. Right: roughness (solid line) on scales comparable to the length $L_s$ of a particle shower will scatter and reduce the coherence of out-going radiation compared to emission through a smooth surface (dashed).}
   \label{fig:rough_sketch}
\end{figure}

The most important effect of the lunar topology on the Askaryan emission from both types of particle cascades is that of the rough lunar surface, as shown in Fig.~\ref{fig:rough_sketch}. On large scales, surface features such as craters and hills will tend to cause cosmic rays to interact pointing into the local surface, so the radiation is directed into the Moon. On small scales, roughness over the area through which the radiation exits the Moon can cause decoherence effects. These effects will be larger for neutrino interactions, where the outgoing radiation transmits through a larger portion of the surface. The exact effects of lunar surface roughness are the least well understood aspect of the theory of lunar Askaryan emission.

In the next section, the basic strategy for detecting these ultra-short signals using a large single-dish telescope such as FAST is described.

\section{Observational Requirements}
\label{sect:observations}

Observing lunar Askaryan pulses with a conventional radio telescope presents a unique challenge. The signal must be sampled at full time resolution, and searched for an excess of impulse-like events. The signal must be de-dispersed in order to compensate for the phase delays induced by the Earth's ionosphere, and lunar pulses must be distinguished from impulsive RFI and thermal noise fluctuations. The more general requirements of an observation are described below, followed by a detailed description of these specialised requirements in their respective subsections.

The principle of searching for lunar Askaryan pulses with a ground-based radio-telescope is relatively simple: point the telescope at the Moon, and search for nanosecond pulses. The coherence of the radiation means that the sensitivity scales linearly with both antenna effective area and the available bandwidth. The relevant frequency range is from approximately $100$~MHz, below which the emission will be too weak to detect, up to approximately $2$~GHz, above which the radiation is emitted over such a narrow angular range that the probability of detection is too low. The proposed ultra-wideband single-pixel feed (SPF; \cite{Li13}), with a potential bandwidth from $270$ MHz to $1.45$~GHz, will thus provide the greatest sensitivity per beam for lunar observations, with bandwidth--area product of order $60,000$ m$^{2}$~GHz, compared to the previous maximum of $\sim1300$ for the first Parkes experiment (\cite{Hankins96}).

As much of the lunar disc (diameter $\sim30^{\prime}$) as possible should be observed simultaneously in order to maximise the event rate, or at least the lunar limb, due to the `limb brightening' effect (Sec.~\ref{sect:emission}). In the GHz regime, the FAST beam however is very small compared to the Moon, which is the main disadvantage of using a single beam from a single large dish for lunar observations. The 19-beam FAST multibeam receiver (\cite{Nan11}) will somewhat alleviate this problem, albeit with a smaller bandwidth of $1.23$--$1.53$~GHz\footnote{The expected receiver bandwidth has now been changed to $1.05$--$1.45$~GHz (D.~Li, private communication).}.

\begin{figure}
   \centering
   \includegraphics[width=\textwidth, angle=0]{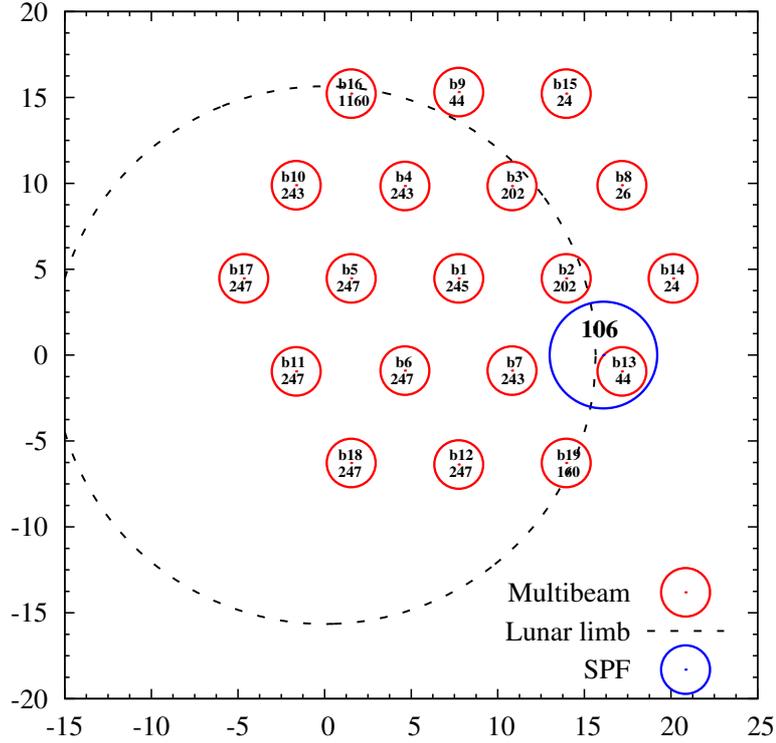}
   \caption{Configuration of simulated FAST beams compared to the Moon, at distance $3.844 \cdot 10^{8}$~m. Red: central positions of the 19 beams of the multi-beam receiver; blue: assumed pointing position of the single-pixel wideband feed. The estimated T$_{\rm sys}$ of each beam is also shown. The beam size is calculated at the geometric mean of the bandwidth, i.e.\ $(f_{\rm min} f_{\rm max})^{0.5}$, while T$_{\rm sys}$ is averaged over the band.}
   \label{fig:multi_moon_point}
   \end{figure}

A possible observing configuration for both the multibeam and SPF is shown in Fig.~\ref{fig:multi_moon_point}. The multibeam configuration allows four on-limb beams (b2, b3, b16, and b19), which will be the most sensitive to Askaryan signals, while the optimal SPF pointing is $0.44^{\prime}$ off-limb. Such a slightly off-limb pointing tends to be advantageous because the noise in each beam will be dominated by the lunar thermal emission, at approximately $225$~K (\cite{troitskijtikhonova70}). Pointing slightly off-limb thus significantly reduces the lunar noise contribution while maintaining sensitivity to the majority of signal events.

The small angular coverage of standard receivers, combined with the dominance of lunar noise, makes a phased-array feed (PAF) well-suited for lunar observations. In the ideal case, a PAF could be built which would simultaneously observe the entire lunar disc over a bandwidth comparable to that of the SPF, without too great an increase in system temperature. While the building of such a receiver would be challenging, it would also be useful for other FAST observation programs, such as pulsar searches.

Regardless of which beam is used, the special requirements of dedispersion, RFI discrimination, and the likely management of a high data rate via triggering, will have to be implemented. These are described below.

\subsection{Dedispersion}

\begin{figure}
   \centering
   \includegraphics[width=0.45\textwidth, angle=0]{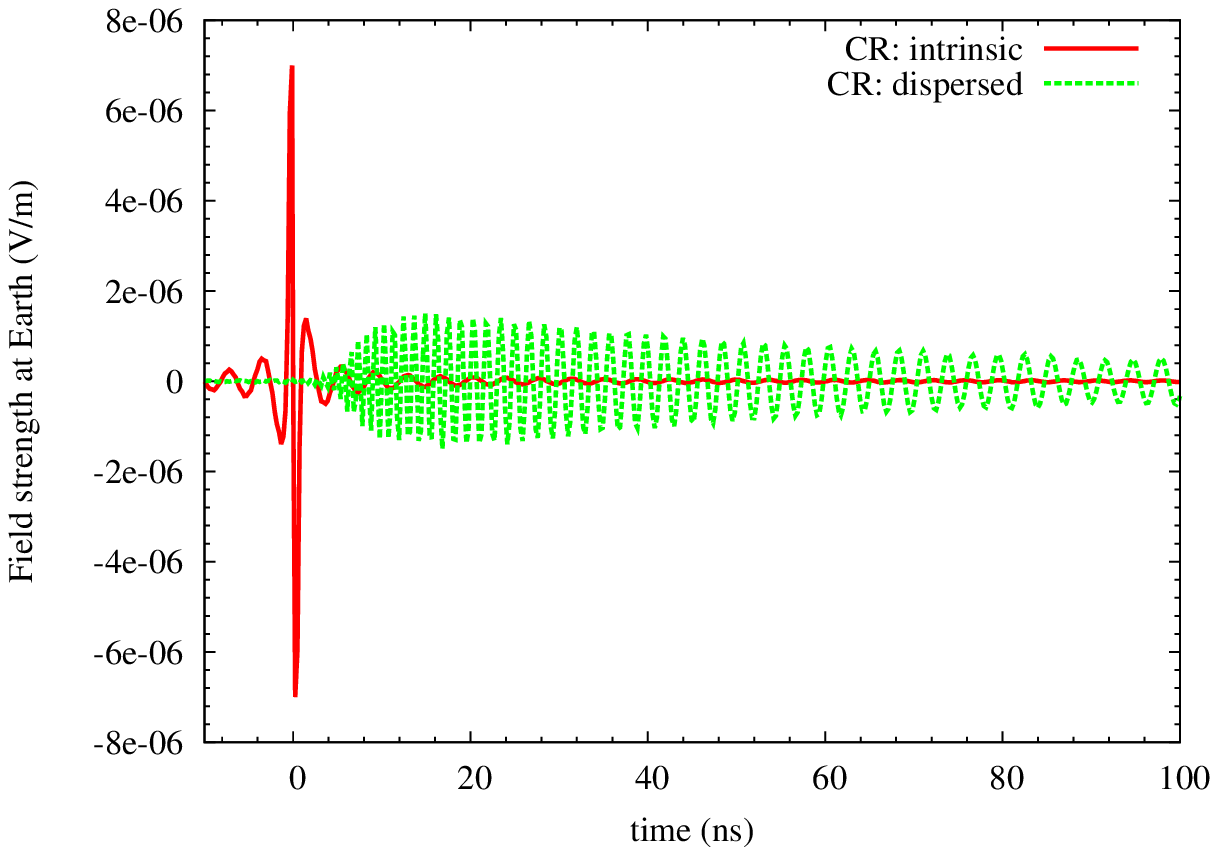} \includegraphics[width=0.4\textwidth, angle=0]{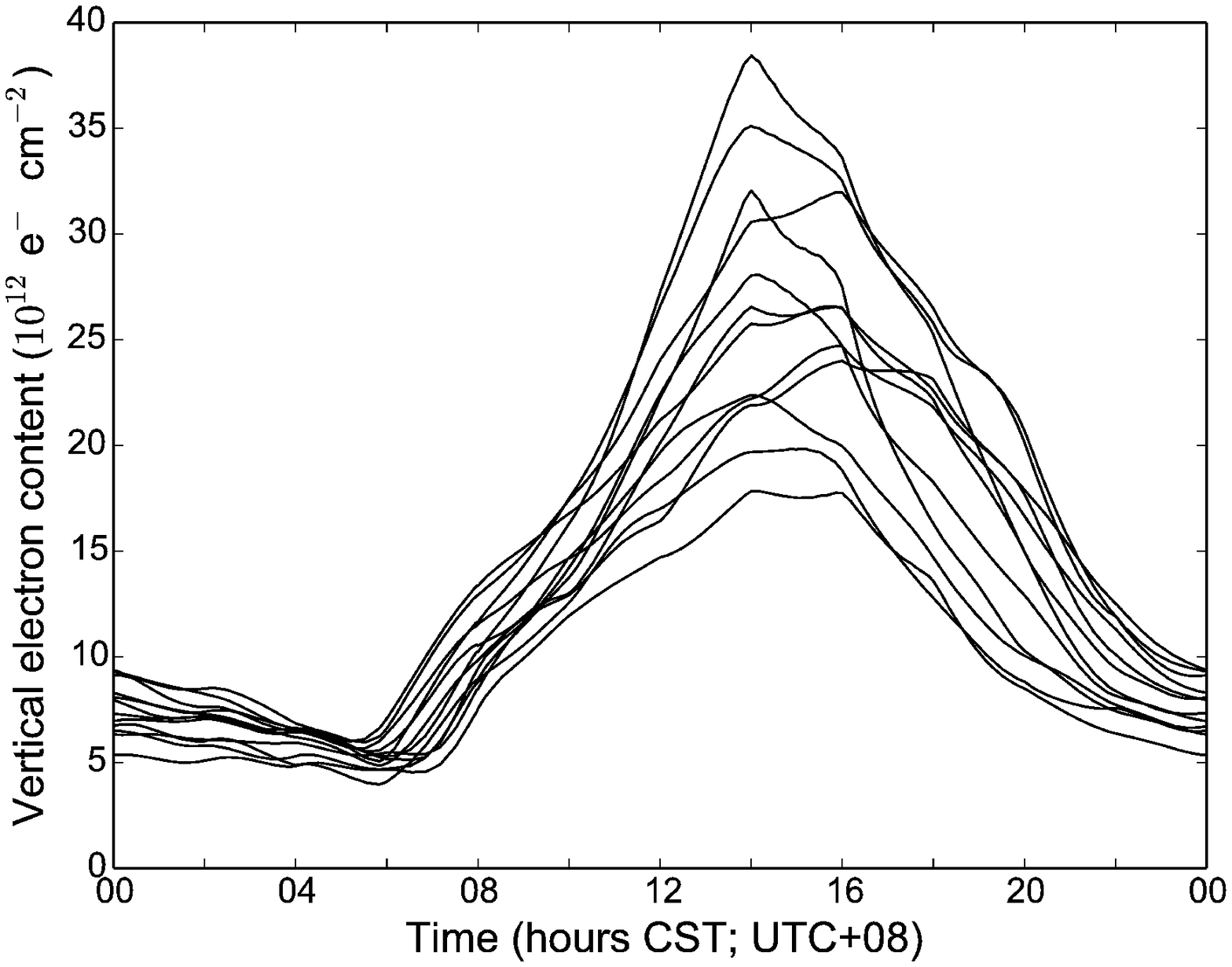}
   \caption{Left: effects of dispersion (green dashed line) on an example cosmic-ray signal (solid red line). The assumed bandwidth is $270$~MHz--$1.43$~GHz, and STEC of $7 \cdot 10^{12}$ e$^-$ cm$^{-2}$. Right: monthly-averaged VTEC values for 2007 at the FAST site as a function of local time (CST), as derived from GPS data (\cite{Noll2010}).}
   \label{fig:dispersion}
\end{figure}

The broad bandwidth and short time-duration of Askaryan pulses means that the dispersion induced by the Earth's ionosphere is significant. Since the main characteristic of the pulses is their short time-duration, any signal smearing due to dispersion over the observation bandwidth will drastically reduce the experimental sensitivity. The dispersion $\Delta t$ induced in a pulse can be calculated via Eq.~\ref{eq:dispersion}:
\begin{eqnarray}
\Delta t {\rm (s)} & = & 
1.34 \cdot 10^{-3} \, {\rm STEC} \, \left(f_{\rm min}^{-2} - f_{\rm max}^{-2} \right) \label{eq:dispersion}
\end{eqnarray}
where $f_{\rm min}$ and $f_{\rm max}$ are the minimum and maximum observation frequencies (Hz), and STEC is the slant total electron content  (e$^-$ cm$^{-2}$) along the line of sight. For a bandwidth of $1.23$ to $1.53$ GHz, and typical night-time STEC of $7 \cdot 10^{12}$ e$^-$ cm$^{-2}$, the pulse is smeared over $\Delta t = 2.2$ ns, which is comparable to the inverse bandwidth. During the day, or during periods of significant solar activity, the dispersion will be greater. The effects over broader bandwidths, and at lower frequencies, will also be greater. 
An example of dispersive effects is given in Fig.~\ref{fig:dispersion}(left), in the case of a $270$~MHz--$1.43$~GHz bandwidth, and STEC of $7 \cdot 10^{12}$. This can be compared to the measured VTEC (vertical TEC, equal to STEC only for a source at zenith) at the FAST site, with monthly averages for $2007$ (during a solar minimum) shown in Fig.~\ref{fig:dispersion}(right). While dispersion can be compensated for in digital signal processing, its accuracy is limited by knowledge of the Earth's ionosphere at the telescope location, so that ionospheric monitoring is vital to the success of the experiment.

The necessary real-time monitoring and correction has been successfully used at the LUNASKA experiment at Parkes (\cite{Bray13}), thus demonstrating the feasibility of this method, albeit only over the $1.23$--$1.53$~GHz band. It may be possible however to use the Faraday rotation of the polarised component of the emission of the Moon itself to determine the dispersion along the line-of-sight (\cite{McFadden12}).

\subsection{RFI Discrimination}

Askaryan pulses appear very similar to bandwidth-limited impulses. Two types of event can imitate the signal: random noise fluctuations, and short-time-duration RFI, both of which must be excluded in order to identify a lunar Askaryan signal. Rejecting the former is simply a matter of setting a sufficiently high detection threshold, although the exact noise statistics can become quite complex (\cite{BrayEkersRoberts}). For radio-telescope arrays, rejecting RFI also becomes relatively easy, since the angular resolution of arrays readily allows signals not of lunar origin to be rejected. For single-dish experiments such as FAST however, RFI discrimination becomes much more difficult. The methods developed by the LUNASKA experiment at Parkes (\cite{Bray13}) will be equally applicable at the FAST telescope, and are discussed below.

RFI that can imitate an Askaryan pulse is necessarily broadband and of short time duration. Unlike the narrow-band RFI environment at radio-telescope facilities, which tends to be well-monitored, sources of impulsive RFI are largely unknown. Examples found in the Parkes experiment include air conditioners in the control room, and noise-calibration diodes located in the telescope receiver. Such pulses will differ from true lunar pulses in the following ways: their origin, their polarisation, their timing, and their lack of an atmospheric dispersion signature. Of these, the signal polarisation does not generally provide a strong discriminant, since the expected polarisation of lunar signals can only be approximately predicted (and at low frequencies, will be Faraday-rotated over the bandwidth). The expected dispersion signature is however a strong discriminant --- terrestrial RFI will obviously not be dispersed, unless it is reflected off an object outside the atmosphere, in which case it will be doubly dispersed. RFI also tends to arrive in bursts, and is usually longer in duration than Askaryan pulses, so that the high time resolution required to search for the pulses also helps discriminate against RFI events.
The strongest discriminant used in the Parkes experiment however was the required lunar origin, using an anti-coincidence veto between different beams on the multibeam receiver.

Lunar-origin pulses are point-like events, and all but the most energetic events will appear in at most one beam of a multi-beam receiver. Terrestrial RFI however will not enter through the main telescope beams, and will tend to appear in multiple beams at once. Events appearing in more than one beam therefore will likely not be lunar in origin. Any beams pointing away from the Moon (such as beams 8, 14, and 15 in Fig.~\ref{fig:multi_moon_point}) will have a lower system temperature, and therefore be especially sensitive to RFI events. Similar methods would also apply to beams formed from the elements of a PAF --- in the case of a single-pixel feed however, no equivalent method could be used, so that RFI discrimination would be much more difficult.

\subsection{Triggering}

The lunar Askaryan technique involves searching for a signal which closely resembles a bandwidth-limited impulse, being only nanoseconds in duration. Therefore, the full time resolution of the telescope must be preserved, since any averaging --- or splitting of the bandwidth --- will dilute the signal power. The bandwidth must also be sampled at high precision, in order to simultaneously determine the rms noise voltage, and provide a sufficient dynamic range for signal detection. For the FAST multibeam, with two polarisation channels per beam, each sampled at $\sim 1.024$~GHz with $8$-bit precision, the data rate would be approximately $40$~GB/s. If this is too high for baseband recording to be practical, a real-time trigger must be implemented to detect likely candidate pulses in real time, and record snapshots of the data. Such a trigger method has been used in all previous searches for lunar pulses, with the exception of the NuMoon observations at Westerbork.

A real-time trigger will inevitably not be as sensitive to lunar pulses as a full offline analysis could be. In the case of a radio-telescope array, the main challenge is to combine information from several antennas in real-time to form a trigger, which is the biggest advantage of a single-dish experiment such as FAST over such instruments. Dedispersion in real-time however can prove difficult, both computationally, and because information on the total electron content along the telescope line-of-sight is often not immediately available. Effects on the signal due to a finite sampling rate, and phase randomisation due to downconversion, may also need to be corrected for (\cite{Bray12}), which again is more difficult in real time.

To compensate for imperfect sensitivity, the threshold of a real-time trigger can be set well below the theoretical signal detection threshold, producing a high rate of candidate events to ensure that any signal event is recorded. The optimal trigger rate depends on the dead-time of the system upon triggering, which in turn depends on the amount of data that must recorded for each trigger. As discussed above, single-dish experiments such as FAST typically have trouble discriminating against terrestrial RFI. While the anti-coincidence method rejects almost all impulsive RFI with high accuracy, it may be necessary to record microseconds' worth of data from all beams upon each trigger in order to help discriminate against the small fraction of RFI that remains.

All the above methods --- dedispersion, RFI discrimination, and triggering --- were used in the LUNASKA experiment at Parkes to reach a sensitivity very close to the thermal noise level (\cite{Bray13}), indicating that this will also be possible at FAST (at least in the case of a multibeam or PAF receiver). We therefore proceed to simulate the sensitivity of a lunar Askaryan experiment at FAST on the assumption that the limiting sensitivity of the thermal noise level can be reached.

\section{Simulations}
\label{sect:simulations}

Simulations of UHE particle interactions in the Moon, and the associated radio-wave production and propagation, were performed by a detailed Monte Carlo code (\cite{JamesProtheroe09a}). The program interacts UHE particles with the Moon, calculates the radiation strength using parameterisations based on Monte Carlo simulations of particle cascades (\cite{ZHS92,AMhadronic,Alvarez-Muniz06}), accounts for transmission through, and absorption in, the rough lunar surface, and determines if the signal is detectable by a simulated instrument. The accuracy of the simulation has been confirmed by analytic calculations (\cite{Analytic}). Note that while the simulation implicitly assumes cosmic-ray primaries to be protons (the parameterisations on which they are based are made with protons), the resulting emission will be highly insensitive to the nature of the primary particle, since the mechanisms producing an excess charge --- and hence Askaryan emission --- in particle cascades become important only below $\sim100$~MeV\cite{ZHS92}, and the total electromagnetic fraction is a weak function of primary nucleon energy\cite{AMhadronic}.

The main uncertainty in simulating the signal from cosmic-ray interactions comes from the interaction of the cosmic rays with lunar surface features, which tend to cause cascades to occur on locally unfavourable slopes. A `worst case' scenario is simulated by forcing all cosmic rays to interact with unfavourable local slopes, as described in James \& Protheroe (\cite{JamesProtheroe09a}), thereby putting a lower bound on the emission, whereas the standard simulation assumes a random surface slope. The reality is expected to be somewhat closer to the `random slope' calculation, since only at very low angles of incidence will large-scale surface features affect the interaction geometry. Therefore, the mildly optimistic method only is treated here.

\subsection{FAST receivers}
\label{sect:receivers}

The sensitivity of FAST is simulated assuming three receivers. The first is the planned 19-receiver multibeam with intra-beam spacing of $6.2^{\prime}$, calculated by scaling the $29.1^{\prime}$ spacing of the Parkes multibeam by $64/300$ (the ratio of telescope diameters). Each beam was modelled as having an LCP/RCP receiver with $T_{\rm sys}=25$~K, pointed at the Moon as shown in Fig.~\ref{fig:multi_moon_point}. Such circularly-polarised receivers can not be aligned to take advantage of the expected radial linear polarisation of Askaryan signals. This particular configuration has not been optimised, but the placement of four beams very close to the lunar limb should be close to optimum. Airy beams appropriate to a 300~m aperture were assumed, with an effective area of $50,000$~m$^2$ per beam (equivalent to $70$\% aperture efficiency). The observation band was $1.23$--$1.53$~GHz, with the contribution of lunar thermal noise integrated for each beam across this band, assuming a lunar blackbody  temperature of $225$~K. The detection criteria assumed a simple voltage threshold trigger requiring a coincidence in both the RCP and LCP channels of a single beam at $5.9$ times that of the thermal noise, for a double-coincidence rate off pure noise of five times per year for all $19$ beams. Perfect correction for ionospheric dispersion and instrumental effects, and perfect discrimination against RFI, is also assumed. While obviously unrealistic for a real-time trigger, such sensitivity is feasible in offline analysis, provided the real-time trigger threshold is set sufficiently low.

The second receiver considered is the proposed broadband single-pixel feed (SPF), operating from $270$~MHz to $1.45$~GHz, but otherwise identical to the multibeam receivers. For this simpler configuration, several pointing positions near the lunar limb were simulated, representing different trade-offs between reduced lunar noise from off-limb pointing, and a corresponding reduction in sensitivity to signal events. It was found that the optimal pointing position was $0.5^{\prime}$ off the lunar limb, which is the pointing position shown in Fig.~\ref{fig:multi_moon_point}.

The third receiver considered is a phased array feed (PAF) of sufficient size to cover the entire lunar surface with an effective area of $50,000$~m$^2$. A base $T_{\rm sys}$ of $50$~K was used, to which was added $225$~K of lunar emission. Dual linear polarisations were assumed for this PAF, with an either-or detection threshold of $8.6$~times the thermal noise in each polarisation channel. Note that the linearly-polarised receivers will interact with the radial linear polarisation of the signal to create a non-radially-symmetric sensitivity about the lunar limb. The ASKAP PAF is sensitive to the frequency range of $700$~MHz--$1.8$~GHz, all of which could  be digitised and used simultaneously in a UHE particle search. Therefore, this is the bandwidth chosen for the simulated PAF on FAST. Sufficiently many beams to cover the entire lunar disc were assumed.

Together, these three options represent a range of trade-offs from high-sensitivity, low-coverage observations (the single-pixel feed), through medium-sensitivity, medium-coverage observations (the multibeam), to low-sensitivity, high-coverage observations (PAF), although the broad bandwidth of the PAF will make it more sensitive to Askaryan pulses than the multibeam. All three were applied to the simulated radio signals from cosmic ray and neutrino interactions in the Moon.

\subsection{Sensitivity to Cosmic Rays}

\begin{figure}
   \centering
   \includegraphics[width=0.6\textwidth, angle=0]{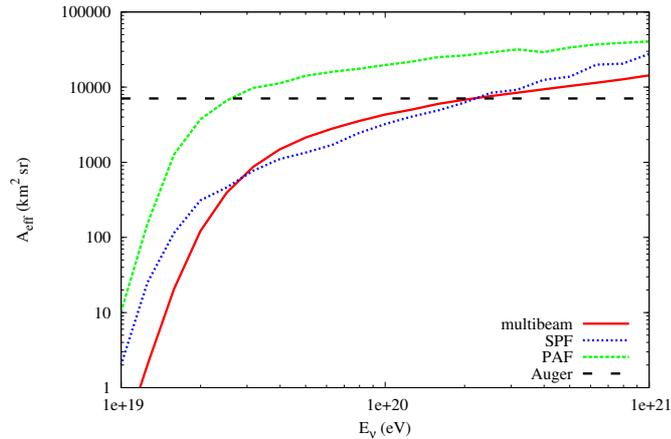}
   \caption{Simulated effective apertures of different FAST receivers ($19$-beam `multibeam', single-pixel feed `SPF', and a phased-array feed `PAF') to cosmic rays. Also shown for comparison is the instantaneous aperture of the Pierre Auger observatory, which is constant in this energy range.}
   \label{fig:cr_app}
\end{figure}

The corresponding effective apertures to cosmic rays from the receivers described in Sec.~\ref{sect:receivers} are shown in Fig.~\ref{fig:cr_app}. The implications of receiver choice are clear: a broadband phased-array feed, able to view the entire Moon with close to full sensitivity, provides a much higher sensitivity over the full cosmic ray energy range than either the SPF or the PAF. While the SPF may have a marginally broader bandwidth and lower system temperature, the ability to view the entire Moon is of paramount importance. It is only above $10^{20}$~eV, where a cosmic-ray signal becomes strong enough to enter through the sidelobes of the SPF, that the effective area is more than $10$\% that of a PAF.
For all receivers simulated, the detection threshold for FAST is approximately $10^{19}$~eV, and thus the telescope should be sensitive to the full energy range at which the cosmic-ray arrival direction contains information.

\begin{figure}
   \centering
   \includegraphics[width=0.6\textwidth, angle=0]{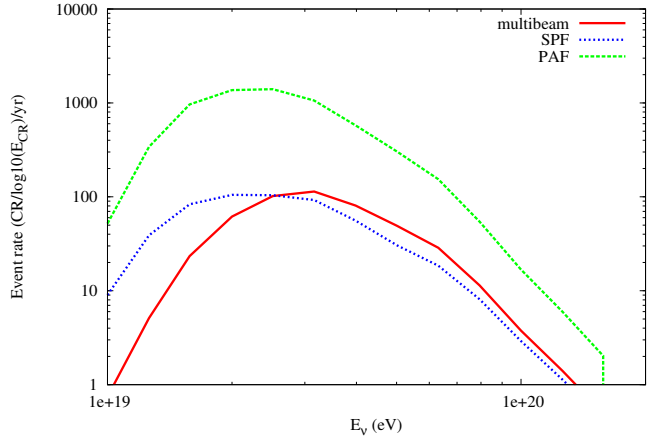}
   \caption{Expected cosmic-ray detection rate for the FAST observations. Values above $2 \cdot 10^{20}$ can not be calculated, because the flux at these energies is unknown.}
   \label{fig:cr_rates}
\end{figure}

In order to determine the event rate, the apertures of Fig.~\ref{fig:cr_app} have been convolved with the approximate cosmic-ray flux measured by the Pierre Auger observatory (\cite{AUGERrio}). The results are shown in Fig.~\ref{fig:cr_rates}. Combining the very steep cosmic-ray spectrum (no harder than $dN/dE \sim E^{-3}$) with the sharp turn-on of the effective experimental aperture above threshold produces a sensitivity which is sharply peaked, and spans only a relatively small energy range. The peak flux is expected between $2$--$3 \cdot 10^{19}$~eV, depending on the receiver. The integrated event rates for the multibeam, SPF, and PAF are $48$, $55$, and $630$ events per full year's worth of observation respectively, while the rates of `interesting' events with energies above $5.6 \cdot 10^{19}$~eV are $2$, $3$, and $39$ per year (the rate for Pierre Auger is approximately $30$ per year).

From the estimates in Fig.~\ref{fig:cr_rates}, a PAF on FAST would detect approximately one event per two six-hour observations; equivalently, a month spent tracking the Moon whenever it is visible should yield of the order of fifteen events. In the extreme energy range above $5.6 \cdot 10^{19}$~eV where cosmic ray arrival directions carry useful information on the source flux, the event rate --- adjusted for a lunar visibility of perhaps $25$\% --- is ten per year. While this rate is not sufficient to compete directly with the Pierre Auger observatory, it raises the prospect of FAST making a first detection using the lunar Askaryan technique, thereby paving the way for instruments such as the SKA to make even-more-sensitive observations.

\begin{figure}
   \centering
   \includegraphics[width=0.6\textwidth, angle=0]{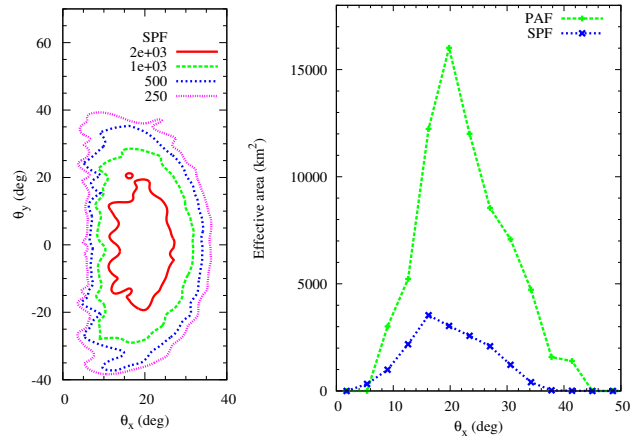}
   \caption{Effective area (km$^2$) to cosmic rays of $E_{\rm CR} = 10^{19.5}$~eV as a function of their arrival direction relative to the Moon.
    Here, $\theta_x$ and $\theta_y$ are angles from the Moon in the $x$ and $y$ directions in the sky plane, defined with the Moon located at $(0,0)$, and the SPF pointing at $(x,y)=(0.27,0)$. Left: contour plot of the effective area of the SPF; Right: one-dimensional comparison of the SPF with the PAF along the $\theta_y=0$ line. Note that the effective area of the PAF is almost radially symmetric due to the uniform coverage of the lunar surface.}
   \label{fig:sky_coverage}
\end{figure}

The total number of cosmic-ray detections is not the only experimental outcome however. The particular geometry of lunar Askaryan emission means that different observation modes will be sensitive to different parts of the sky (\cite{JamesProtheroe09b}). The instantaneous effective area to cosmic rays of $5 \cdot 10^{19}$~eV of both the FAST SPF and PAF is shown in Fig.~\ref{fig:sky_coverage} as a function of their arrival direction relative to the Moon. In the case of the PAF, which is assumed to view the entire Moon, the coverage is radially symmetric about the lunar centre, and is distributed in a broad band between approximately $15^{\circ}$ and $40^{\circ}$ from the Moon. The coverage of the SPF is maximal in a region of approximate dimensions $30^{\circ} \times 70^{\circ}$, centred $\sim20^{\circ}$ from the Moon in the direction given by the beam-pointing position on the lunar limb. Thus these two modes have not only different energy-dependencies of their sensitivity --- they also have different angular-dependencies as well. Note that the directional sensitivity dependence means that by scheduling observations when the Moon is in a part of the sky near a candidate UHE particle source (such as Centaurus A), the low fraction of telescope observation time compared to the $100$\% duty cycle of an experiment such as the Pierre Auger Observatory can be somewhat compensated for (\cite{James11}).

\subsection{Sensitivity to Neutrinos}

\begin{figure}
   \centering
   \includegraphics[width=0.6\textwidth, angle=0]{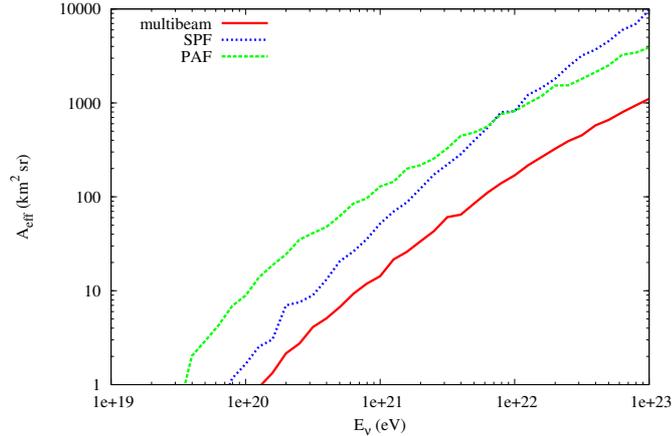}
   \caption{Simulated effective apertures of different FAST receivers ($19$-beam `multibeam', single-pixel feed `SPF', and a phased-array feed `PAF') to UHE neutrinos.}
   \label{fig:nu_app}
\end{figure}

The effective aperture to UHE neutrinos of the three receiver packages is given in Fig.~\ref{fig:nu_app}. The shape is quite different to that of cosmic rays, due to the different interaction phenomenology of the particles. Since the majority of neutrinos interact too deeply for their radiation to escape the surface, and only a fraction of their energy is given to hadronic cascades, the total effective aperture of the simulated FAST receivers to UHE neutrinos is much lower than to UHE cosmic rays. This is why the aperture continues to increase with energy: particles with higher energy can be detected at greater depths, and when giving a lower fraction of their energy to hadronic cascades. Thus at extreme energies above $10^{22}$~eV, the SPF is more sensitive to UHE neutrinos than the PAF, since its lower system temperature, and lower minimum frequency, allow it to probe more deeply into the Moon. Additionally, since neutrinos can penetrate a significant fraction of the lunar limb, and thus undergo interactions that point out of the surface, it becomes possible to detect the peak emission at high frequencies where the emission cone is relatively narrow. Hence, the PAF has a greater sensitivity than the SPF to particles at `low' energies below $10^{21}$~eV. Given the spectral downturn in the cosmic ray spectrum near $10^{19.5}$~eV, this suggests the PAF as the optimum receiver for UHE neutrino searches as well as for UHE cosmic rays.

While no neutrino in the UHE energy range has been detected (those observed by IceCube had interaction energies of at most $10^{15}$~eV (\cite{IceCube}), which is well below the energy range considered here), several experiments have already searched for them, and placed limits on their flux. As well as the Pierre Auger experiment, two radio-detection experiments have used the Antarctic ice sheet as an interaction medium to search for Askaryan pulses from UHE neutrinos. These were ANITA (\cite{ANITA}), which consisted of an array of broadband receivers mounted on a high-altitude balloon, allowing a very large volume of ice to be observed; and RICE (\cite{RICE}), which was an array of radio receivers embedded in the ice sheet itself.

\begin{figure}
   \centering
   \includegraphics[width=0.6\textwidth, angle=0]{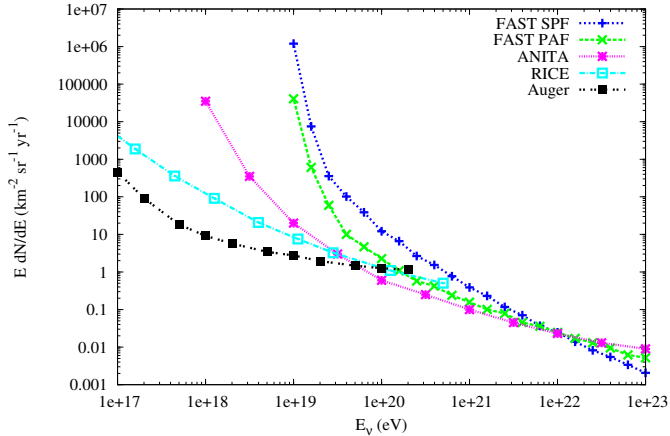}
   \caption{Limits on the UHE neutrino flux which would arise from a $1000$ hr observation campaign with FAST, using the PAF or SPF, compared to existing limits from the ANITA (\cite{ANITA}), RICE (\cite{RICE}), and Pierre Auger (\cite{AUGERrio}) experiments.}
   \label{fig:nu_lim}
\end{figure}

The limits on the UHE neutrino flux that would result from a dedicated $1000$-hr observation campaign with the FAST PAF or SPF are compared to existing experimental limits in Fig.~\ref{fig:nu_lim}. While the limits would be competitive in the highest-energy range with those of experiments such as RICE and ANITA, a much longer observation campaign would be required in order to improve upon them. UHE neutrino detection would also necessarily involve discriminating against UHE cosmic ray events, which would be detected more commonly.

\section{Discussion}
\label{sect:discussion}

The estimated sensitivities of UHE particle experiments with FAST calculated in Sec.~\ref{sect:simulations} are significantly higher than any previous experimental sensitivity using the lunar Askaryan technique. The achievable limits on a UHE neutrino flux show no significant improvement upon those from the best existing experiments, in particular ANITA and RICE, which were specialised experiments designed specifically to search for UHE neutrinos. The expected rate of cosmic ray detections, once the fraction of time at which the Moon will be visible to FAST is taken into account,
 is lower than that of the existing $3000$~km$^2$ Pierre Auger cosmic ray observatory. This makes it clear the FAST will not be a long-term observatory for UHE particles.

The above results however do point towards a very clear goal of FAST: to make the first detection of an UHE particle using the lunar Askaryan technique. The outcome of such a campaign should be to detect `a few' lunar UHE particle events, regardless of their nature, and thereby demonstrate that the lunar Askaryan technique is feasible. This could be achieved with the known cosmic ray flux using a broadband PAF in as little as one week's worth of observations (seven six-hr periods). For the SPF or the multibeam, a much longer campaign of several months' duration would be required. Such a detection would be a major milestone in both astroparticle physics and radio astronomy in general, and of course the FAST telescope itself, and pave the way for even more sensitive observations with the Square Kilometre Array\cite{SKA}.

The major limits to the sensitivity of such observations will be the effects of surface roughness, the ability to reject RFI, and whether or not a broadband phased-array feed is built for FAST.

The exact effects of lunar surface roughness on the radio signal from UHE particle cascades below the surface are not yet known. This is particularly important for cosmic-ray detection, since the initial interactions of these particles are sensitive to large-scale features. Recent work has shown that it is possible to model the effects of small-scale roughness on the propagation of coherent radio pulses from particle cascades (\cite{JamesArena2012}), while detailed maps of the large-scale lunar surface are also available. It is vital that this work be continued, and this source of uncertainty be removed.

RFI rejection, which must be performed with high fidelity in order to identify lunar Askaryan pulses, is much more difficult on a single antenna that a telescope array. Numerous methods exist to do this, as discussed in Sec.~\ref{sect:observations}. However, the degree of signal purity required to confirm a detection is somewhat higher than that to set a limit. Experience has also shown that it is difficult to anticipate the nature of the nanosecond-scale RFI environment. The successful methods employed to reject RFI by the LUNASKA collaboration in their recent experiment at Parkes (\cite{Bray13}) indicate that complete RFI rejection is indeed possible with a multibeam receiver, and hence also with a phased-array feed. It is unlikely however that a single-pixel feed would be able to achieve this goal.

The building of a broadband PAF with capabilities matching those modelled ($700$~MHz to $1.8$~GHz bandwidth) will pose a significant challenge. In particular, covering the entire Moon to half-power sensitivity at $1.8$~GHz (beam width $2^{\prime}$) would require over $200$ beams to be formed. Since most of the signal will originate from the lunar limb due to the limb-brightening effects, forming beams to cover the limb would capture most of the signal --- additionally, the system noise for these beams would be reduced. A preliminary simulation of such a configuration indicates that when using $24$--$36$~beams (spacing of HPBW at $870$~MHz and $1.3$~GHz respectively), the reduced lunar noise compensates completely for the imperfect lunar coverage. Therefore, the highly promising simulation results for the phased-array feed may indeed be achievable in practise.

\section{Conclusion}

We have described how FAST, using the lunar Askaryan technique to search for nanosecond pulses of radiation, could detect the flux of ultra-high-energy cosmic rays, and potentially ultra-high-energy neutrinos. Using the planned L-band multibeam receiver, the known cosmic ray flux could be detected in a few months' of observation time, making FAST the first such telescope to do so. A phased-array feed capable of forming $24$ beams over a $700$~MHz to $1.8$~GHz bandwidth would make FAST a far superior instrument for such observations, allowing the flux to be detected in as little as a week.

Using the lunar Askaryan technique with FAST would be technically challenging, and very different from the usual telescope observation modes. The techniques of signal detection and RFI rejection developed for the LUNASKA observations with the Parkes multibeam have demonstrated that such observations are possible however with a single large antenna, allowing the theoretical sensitivity limit of telescope system temperature to be reached. On the theoretical side, advances in the understanding of the effects of lunar surface roughness are required in order to reduce the uncertainty in the event rate, and improved estimates of the achievable angular and energy resolution should be made, although the expectation is that these will not achieve the accuracy of experiments such as the PAO. 

The promise of the lunar Askaryan technique with FAST is such that both technical and theoretical efforts would be well-justified, and would allow FAST to become the first terrestrial telescope to successfully utilise the lunar Askaryan technique.

\label{lastpage}

\end{document}